# MOBILE DEVICE USERS' SUSCEPTIBILITY TO PHISHING ATTACKS

F. Ley Sylvester

Department of Computer Science, Tarrant County College, Fort Worth, Texas, USA

## ABSTRACT

*The mobile device is one of the fasted growing technologies that is widely used in a diversifying sector. Mobile devices are used for everyday life, such as personal information exchange – chatting, email, shopping, and mobile banking, contributing to information security threats. Users' behavior can influence information security threats. More research is needed to understand users' threat avoidance behavior and motivation. Using Technology threat avoidance theory (TTAT), this study assessed factors that influenced mobile device users' threat avoidance motivations and behaviors as it relates to phishing attacks. From the data collected from 137 mobile device users using a questionnaire, the findings indicate that (1) mobile device users' perceived susceptibility and severity of phishing attacks have a significant correlation with a users' perception of the threat; (2) mobile device users' motivation to avoid a threat is correlated to a users' behavior in avoiding threat; and (3) a mobile device user's susceptibility to phishing attacks can be reduced by their perception of the threat. These findings reveal that a user's perception of threat increases if they perceive that the consequence of such threat to their mobile devices will be severe, thereby increasing a user's motivation and behavior to avoid phishing attack threats. This study is beneficial to mobile device users in personal and organizational settings.*

## 1. INTRODUCTION

Mobile devices now perform as personal computers instead of a communication device as technological and computing capabilities increase. Mobile devices are used to perform everyday interactions and transactions and have contributed to increased security concerns for users. Security threats are increasing as the global population continues to adopt mobile device use. Phishing attackers continue to look for ways to penetrate mobile devices [1] [2]. As users' dependence on mobile devices increases, so is their susceptibility to information technology threats. Therefore, it has become necessary to understand the avoidance motivation and behavior of users. It is important to understand mobile device user behaviors as their dependency on mobile devices increases [3].

Mobile device users are more susceptible to phishing attacks than desktop users [4]. Some mobile device users are not aware of phishing attack techniques and may not realize that they are victims or could become victims of an attack [5] [6]. [7] pointed out the differences in how users interact with computers compared with mobile devices impact vulnerabilities. Users can perform only a subset of activities on their mobile devices and, due to the portable size of their mobile devices, may miss some vital details and click on or open malicious emails. For example, mobile device users may more frequently focus on urgency cues in email and omit unconventional grammar or spelling in emails than computer users, thus increasing the possibility of being a victim of phishing attacks [8]. A change in mobile device users' behavior can reduce the victimization of phishing attacks [9]. Despite the recent increase in attention on IT security driven by the pervasiveness of technology in human interactions, there is a need for research on the impact human behavior has on cybercrime vulnerabilities in the context of mobile devices [10].

Even though mobile devices are one of the fastest-growing markets in the technology sector, there has been limited research done to understand why mobile device users are increasingly falling victims to phishing attacks. Thus, it is essential to understand mobile device users' perspectives on avoiding phishing attacks and how they might be protected from internet security threats while using their devices. Incorporating human behavior into our understanding of mobile device vulnerabilities will aid in the development of the next generation of mobile device security tools and strategies.

Recent research has detailed the growing threat to mobile device security [11], [12]. However, most of these studies focus on mobile vulnerability analysis to internet threats [13], online security tools to thwart IT threats, and malware and phishing attacks concerning websites or URLs [7]. Malware and phishing studies on mobile devices focus on the growing trend of sophisticated malware [14] and techniques to detect phishing attacks and countermeasures [15]; Security studies of mobile applications focus on areas such as spoofing with the use of malicious applications [12]. Several studies note that the increased dependence on mobile devices is concurrent with an increase in users' vulnerability to phishing attacks [16], [3]. Studies on online security tools focus on protecting users from IT threats [17], while a few studies combine mobile device users' avoidance behavior and their susceptibility to phishing attacks.

Although many studies indicate that internet security is a significant concern due to the increased use of mobile devices, there is little available information about mobile device users' perception of phishing attacks as an IT threat. Furthermore, research has not adequately addressed mobile devices users' security behavior in responding to phishing attacks [10]. It is necessary to understand why mobile device users are increasingly falling victim to phishing attacks or online threats [18]. This study examined the factors that make mobile device users susceptible to phishing attacks [7], [3], [4], user avoidance behavior, and motivation to avoid IT threats. In doing so, this study contributes to the body of research used to inform and update mobile device security practices.

## 1.1. The Present Study

The increased reliance on mobile devices can lead to an increased frequency of phishing attacks [19]. Exploring mobile device users' susceptibility to phishing attacks provides scholars and mobile device users with a better understanding of why preventive measures by themselves do not adequately protect against internet attacks. Such insights could help reduce the risky behavior of mobile device users and thereby thwart phishing attacks.

The study intended to investigate mobile device users' susceptibility to phishing attacks and provide insights into reducing users' risky behavior. This study aimed to inform scholars and individuals on a better understanding of internet threat behavior among mobile device users and preventive actions utilizing the technology threat avoidance theory (TTAT). TTAT explores the influence of perceived severity, perceived susceptibility, perceived threat, safeguard effectiveness, safeguard cost, self-efficacy, avoidance motivation, and avoidance behavior [10] on mobile device users' security behavior related to phishing attacks. By using the constructs outlined in TTAT [10], the research presented in this study addresses some of the gaps reported in the literature concerning IT threats. A better understanding of mobile device users' perception and behavior of the phishing attack might lead to a better approach in thwarting phishing attacks. Mobile devices have many of the same capabilities as a laptop or personal computers and are therefore vulnerable to computer security threats. These computer threats are increasing in sophistication and complexity, regardless of the targeted device [18]. The results of this research could help users and practitioners move closer to their shared goal of decreasing mobile device threat vulnerabilities.

The remainder of the paper is organized into five remaining sections. Section 2 provides a comprehensive literature review, including a theoretical foundation and review of related studies. In Section 3, the study details the research methodology, population and sample, research design, instruments, and demographic characteristics used in this study. Section 4 provides an in-depth analysis of the data and the results. Research discussion, limitations, implications, and further research direction are discussed in Section 5; lastly, Section 6 details the conclusion of this study.

## 2. THEORETICAL BACKGROUND AND RELATED STUDIES

Studies on human perspective of security practices on mobile devices and information security have used PMT and TPB to gain a better understanding of what motivates users to comply with information security practices [20], [21]. [22] PMT was PMT to understand internet users' security behavior and mobile device users' intent to take protective measures. [23] studied users' behavior in fighting identity theft by drawing on coping behavior theory, PMT, and TTAT frameworks incorporated the coping behavior theory. [24] studied individual security behavior using a combination of TTAT and PMT.

The theory of planned behavior, protection motivation theory [25], and technology threat avoidance theory [11] have all been used by researchers in the form of behavioral intentions [26], [27] to improve and encourage internet security among mobile device users. Researchers continue to seek additional studies in information security behaviors among individuals [25]. The theories have been used to understand how human behavior can mitigate against IT threats by examining behavior intents and impacts to the security threat.

### 2.1. Technology Threat Avoidance Theory

The technology threat avoidance theory (TTAT) was created by [28] to explain how home computer users made decisions about how to avoid IT threats [10]. It has been used to examine how users can avoid IT threats by using a given safeguard measure [29]. The TTAT was designed to determine factors that influence users to take preventative actions against IT threats [30]. [10] conducted a study using the research model derived from the TTAT to understand how personal computer users avoid IT threats. This theory has been used by [11] and [31] to educate and enhance users' behavior against phishing attacks; it is an essential tool to explain an individual's behavior to avoid malicious attacks [32]. Researchers have examined the eight components of the TTAT (perceived susceptibility, perceived severity, perceived threat, safeguard effectiveness, safeguard cost, self-efficacy, avoidance motivation, and avoidance behavior) as factors that influence how users approach the reality of IT threats.

Attackers use deceptive techniques to carry out fraudulent activities; thus, phishing attacks have increased tremendously [17]. Phishing attacks are methods cybercriminals use to persuade users to provide them with sensitive information [16], [33]; [4] and could be referred to as online identity theft [11]. Researchers have investigated phishing attacks on various platforms, including mobile devices, and they have recorded that these costly attacks are avoidable [11] [34], [35].

## 2.2. Mobile Device and Phishing Attacks

Mobile devices, including smartphones and tablets, have become part of our everyday lives [36], [37]. Over 58% of Americans own a smartphone [38], and 7% of the United States population access the internet through their mobile device [38]. In 2014, 23.53% of mobile devices accessed over 15 billion websites, an increase of 10% from the previous year [36]. The Pew Center reported that over 60% of adult internet users use a mobile device [3].

As people increasingly use mobile devices for their social media and email needs [39], personal data and information are more vulnerable than ever; they must be effectively managed [40]. Mobile device technology has advanced dramatically, and their hardware is comparable to laptops [41]. The use of mobile devices in place of laptops has become more popular. Technology advances such as the faster internet have contributed to increased dependence on mobile devices [42]. However, the increased use of mobile devices and constant internet access have made mobile devices targets for attacks [43]. Although people are more dependent on mobile devices, they are not aware of mobile device security vulnerabilities [41].

The increasing security problems experienced by mobile devices and limited mobile device technology have drawn the attention of researchers [7]. Individuals are not taking appropriate security precautions in protecting their mobile devices [1]. One of the most significant attacks on mobile devices security is malware, a form of a phishing attack [36]. Anti-malware software is available for mobile devices; however, users do not adequately adopt this software [36]. Some individuals use password authentication to secure their mobile devices, but passwords are vulnerable to attack due to their weakness [40], [39]. A better method for mobile device security is to create a pattern that uses data-rich interactions based on a user's interactions with the mobile device and biometrics [40].

While security features and software intended to stop security attacks are continuously updated, cybercriminals follow the technology trends and, as a result, create more sophisticated malware [44]. This evolving landscape is further complicated by user-downloaded applications that can house embedded malware [38], [45]. Information security consists of more than technical issues and frequently depends on a users' attitude towards security [46]. Therefore, it is important to study other methods that users can implement to protect their mobile devices.

## 2.3. Studies on Phishing Attacks

Phishing is a form of identity theft used to gain unauthorized access to personal information [47]. Over the years, researchers have studied if behavioral factors have impacted phishing attacks. Seminal work [48] posited that phishing attacks are growing, and their perpetrators have gained success from it. Their study concluded that implementing anti-phishing tools such as using filtering to detect a spam email, scanning for alerts, monitoring, and providing analysis on an individual's security behaviors will combat phishing.

Not all researchers have attributed a successful phishing attack to an individual's security behavior. However, they recommended the implementation of a security program that combats phishing [49]. [49] educating users on signalling phishing emails is not a solution to stopping phishing attacks. However, other researchers maintain that promoting security awareness among users can reduce phishing attacks without the need for security software [50]. [49] concluded that having a security program, educating users, and maintaining anti-phishing groups would probably help deal with the problem of phishing attacks.

[51] studied what makes individuals vulnerable to phishing using the Routine Activity Theory (RAT) framework. Individuals who had incurred a financial loss after giving their financial details to a cybercriminal were participants in the study. [51] was to examine phishing victimization and the likelihood of phishing attacks based on an individual's computer, internet, and internet banking experience. [52] conducted a study on how to adequately protect against phishing attacks. They emphasized that it is essential to distinguish the attack levels to evaluate the protection techniques [52]. Cybercriminals use computer worms to perform illegal activities such as spamming and redirecting a user's web request, thereby subjecting them to phishing [53]. [52] denote that redirecting a user's website request to obtain personal information is the most popular form of attack. Their study concluded that the browser-side technique is simple to use, effective, and the easiest in protecting users against phishing attacks. [53] posited that the easiest way to protect users against attacks is to remove the perpetrators' motivation in developing worms.

As researchers continue to examine the nature of phishing attacks and how to mitigate them, some have begun to investigate phishing attacks on alternative platforms such as mobile devices [29], [3]. These studies have created various tools and solutions to thwart phishing attacks [11] [54]. Despite the security tools, social engineering attacks continue to be on the rise. [54] posited that individuals should not be completely reliant on technical tools because they do not give complete protection against phishing attacks.

[17] researched phishing awareness; the study asked whether a simulated phishing attack with embedded training was effective in training users on resistance towards phishing attacks. The finding showed that users could adopt a positive email behavior based on a simulated phishing exercise. Therefore, a simulated phishing exercise with embedded training could increase employees' resistance to phishing attacks and help protect the organization's security system. The study concluded that it is essential for employees to know how to protect themselves against phishing attacks. Their study also shows that the simulation of phishing exercises with embedded training can help protect the organization's security system.

A study by [55] gives a theoretical insight into cybersecurity and persistent security behavior among people. There are security measures that individuals need to adhere to; however, some seem to be cumbersome to a subset of people. The complaint includes remembering too many passwords, complex passwords, and long passwords [55]. The authors looked closely into the psychology of people and security behaviors. The study illustrated how a person's perception of security and that of computer security are not aligned, sometimes resulting in a compromised password. The authors used a method that was previously used by [56] to show that about half of the participants seldom care about internet security. The predictive model predicted behavior, indicating that persons that attended computer security training and were continuously reminded were more aware of computer security and more likely to practice it. Half of the 65% respondents wrote passwords somewhere, 80% changed their passwords, and 58% stated that they would not have changed their passwords without the prompts. Together, these results indicated that cybercriminals had taken advantage of user behaviors.

[34] investigated victims of cybercrime with data from a cybercrime victim survey and the RAT to provide a theoretical perspective that explained how technological development had increased cybercrime. Based on this theory, the risk of victimization is affected by various factors. The author described how the digital payment system plays an essential role in cybercrime, primarily due to the increased utilization of online banking and e-commerce. [34] contributed to awareness about phishing victimization by providing insight into phishing victimization protection. The article drew a hypothesis for four elements related to phishing victimization: value, inertia, visibility, and accessibility. The results of the analysis were divided based on the elements.

However, there was an exclusion of inertia from the study due to a lack of validity. The result demonstrated that personal characteristics did not contribute to phishing victimization. The finding showed that economic characteristics, increased online activities, and accessibility did not increase phishing risk. The author concluded that gaining access to a victim's account cannot be stopped; however, proper monitoring of accounts and stopping suspicious activities can prevent the offender from transferring money out of a victim's account. In contrast, the analysis did not recognize people with an increased chance of being a victim.

[6] conducted a literature review on phishing awareness and phishing countermeasures based on the literature reviews of doctoral theses from various universities. [6] found that educating a user about phishing and the implementation of proper phishing applications can be used to protect individuals against phishing attacks. The study identified trends in phishing and suggested ways an individual could use to protect themselves from phishing. The results of the study showed that email filters could be used to reduce phishing significantly. Educating users about identity theft and describing vulnerability, along with implementing proper applications, could help protect mobile device users against phishing. Different approaches were proposed to prevent phishing attacks in the study, but no one approach could stop phishing attacks.

[9] used a qualitative study to examine how social engineering attributes to individual behavior. They asked participants 20 questions to examine their responses to social engineering. The research showed that 42% of network administrators were willing to provide their passwords, and 53% were willing to let others use their email accounts. The attitude of people could impact their perspective on security. [9] found that some people are victims of phishing attacks because of their niceness. [9] pointed out that human resources management also contributed to system security flaws based on their inability to select the appropriate staff. As a result, the apathy of employees in the organization may witness events that attack the organization's security but not report such events. Their findings show that people do not manage their accounts and password correctly, and they could be victims of social engineering.

## 3. RESEARCH METHODOLOGY

This study utilized the technology threat avoidance theory (TTAT; [10]) to test the correlation between perceived severity, perceived susceptibility, perceived threat, safeguard effectiveness, safeguard cost, self-efficacy, avoidance motivation, avoidance behavior, and the mobile device users' security behavior as they relate to phishing attacks. The study addresses a new area of research on what motivates mobile device users to avoid phishing attacks [7]. There has been significant scholarly interest in the correlation between mobile device use and attacks [57]. The results of this study provide scholars and practitioners a new angle on understanding how the behavior of mobile device users impacts the success and frequency of phishing attacks [7].

### 3.1. Research Question and Hypotheses

This study examined, to what extent, if any, do the eight elements of the TTAT (perceived severity, perceived susceptibility, perceived threat, safeguard effectiveness, safeguard cost, self-efficacy, avoidance motivation, and avoidance behavior) influence mobile device users' susceptibility to phishing attacks. Eight research questions (RQ) directly follow:

1. To what extent does perceived severity influence mobile device users' susceptibility to phishing attacks?
2. To what extent does perceived susceptibility influence mobile device users' susceptibility to phishing attacks?

3. To what extent does perceived threat influence mobile device users' susceptibility to phishing attacks?
4. To what extent does safeguard effectiveness influence mobile device users' susceptibility to phishing attacks?
5. To what extent does safeguard cost influence mobile device users' susceptibility to phishing attacks?
6. To what extent does self-efficacy influence mobile device users' susceptibility to phishing attacks?
7. To what extent does avoidance motivation influence mobile device users' susceptibility to phishing attacks? and
8. To what extent does avoidance behavior influence mobile device users' susceptibility to phishing attacks?

These eight research questions were used to generate nine working hypotheses based on the TTAT constructs.

Hypothesis 1a. Mobile device users perceived susceptibility of being attacked by phishing has a positive relationship with their perception of threats.

Hypothesis 1b. Mobile device users perceived severity of being attacked by phishing has a positive effect on perceived threat.

Hypothesis 1c. Mobile device users perceived susceptibility and perceived severity of a phishing attack have a positive interaction effect on perceived threat.

Hypothesis 2. Perceived threat from mobile device users positively affects user's avoidance motivation.

Hypothesis 3: Safeguard effectiveness against phishing attacks positively affects avoidance motivation.

Hypothesis 3a: Perceived threat of phishing attacks and safeguard effectiveness against phishing have a negative interaction effect on avoidance motivation.

Hypothesis 4: Safeguard cost against phishing attacks negatively affects avoidance motivation.

Hypothesis 5: Self-efficacy for taking safeguard measures against phishing attacks positively affects avoidance motivation.

Hypothesis 6: Avoidance motivation positively affects the avoidance behavior of using safeguard measures.

## 3.2. Population and Sample

The population for this study was mobile device users who own a mobile device and are age 18 and older living within the United States. Surveys were administered online and resulted in a demographically diverse sample population, an expected result due to the extensive use of mobile devices in the United States. The Pew Center reported that over 60% of adults use a mobile device to access the internet, and 15% of young adults ages 18-29 rely heavily on their smartphones for online access (as cited in [3]). There is a possibility that participants had limited or no knowledge about phishing attacks, which would yield a range of variance in the TTAT constructs measured by the survey, and this variance will be used to detect moderation effects.

The demographic for this study is diverse; participants include males and females who were 18 years and older, accessed the internet with their mobile devices, and completed a web-based survey.

### 3.2.1. Size and Power

This study examined the data for incomplete surveys, missing survey responses, and outliers. The total participants were N = 137, and eight cases were eliminated due to missing data leaving N = 129. This study used regression analysis, and assumption testing was performed to ensure that regression analysis was appropriate. For this study, Hypothesis 1a, Hypothesis 1b, Hypothesis 2, Hypothesis 3, Hypothesis 4, Hypothesis 5, and Hypothesis 6 were tested using simple linear regression, while Hypothesis 1c and Hypothesis 3a were tested using multiple linear regression.

The minimum sample size for this study was N = 107, using G*Power 3.1.9.2, a power analysis program used in research studies including behavioral studies [58]. The power analysis confirmed 107 samples is appropriate for the study where alpha = .05, power (1 - β err prob) = .95, and effect size = .15. However, QuestionPro collected a random sample size of 137. There were eight missing data, thus, removed from the sample collected. After removing the eight missing data and six outliers, the sample size was N = 123.

### 3.2.2. Demographic Description

Participants for this survey were asked to provide their age group, gender, and education level. Of the 129 participants that completed the survey, 27.13% were between the ages of 18 and 25, 45.74% were between the ages of 26 and 40, 9.30% were between the ages of 41 and 55, and 17.83% were 56 and older. The highest participants were in the 26-40 age group with N = 59 (45.74%), followed by the 18-25 age group represented by N = 35 (27.13%), as shown in Figure 1.

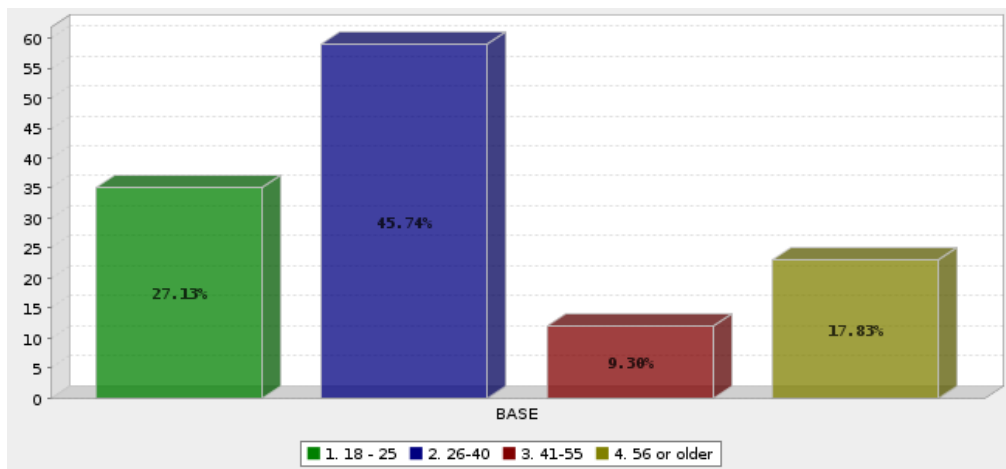

Figure 1. Participants' age group

Figure 2 shows that most participants were females with N = 88 (68.22%). The male participants were at least 50% less than the female participants. The male participants were N = 41 (31.78%) of the sample.

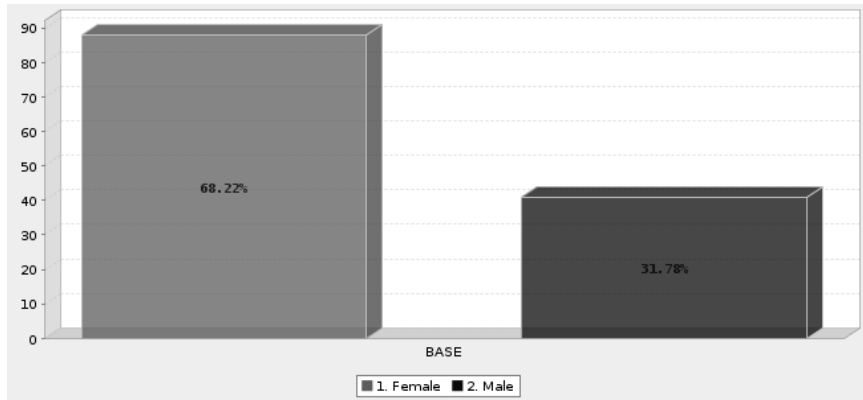

Figure 2. Gender of participants' sample

Participants of the study have attained some level of education. The education level of the participants that completed the survey includes a more significant number of those with some college level of education, with N = 44 (34.11%), followed by those with a high school or equivalent education level, with N = 41 (31.78%). Table 1 reflects the analysis of the education level for the participant.

Table 1. Frequencies of Participants by Education Level.

| Education Level | Frequency | Percentage |
|---|---|---|
| High school or equivalent | 41 | 31.78% |
| Vocational/technical school | 6 | 4.65% |
| Some college | 44 | 34.11% |
| Bachelor's degree | 25 | 19.38% |
| Graduate school | 13 | 10.08% |
| Other | 0 | 0.0% |

## 3.3. Measures

The instrument used for this study was a survey based upon the constructs of the technology threat avoidance theory (TTAT). The instrument contained 44 questions that mapped to the TTAT constructs – perceived susceptibility, perceived severity, perceived threat, safeguard effectiveness, safeguard cost, self-efficacy, avoidance motivation, and avoidance behavior. There were six sections of the survey anchored on 7-point Likert-type scales (1 = strongly disagree, 7 = strongly agree), a seventh section of the survey anchored on a 7-point Likert-type scale (1 = innocuous, 7 = extremely devastating), and the eighth section anchored on 10-point Likert-type scale (1 = not at all confident, 10 = totally confident).

Validity is essential to any measuring instrument used to collect data [59]. An instrument is valid if it measures what it intends to measure and if the measurement is accurate [60]. Data was collected appropriately using an online questionnaire. The method of data collection is relevant to the validity of the research. According to [61], to achieve p < .05, some researchers increase the datasets using various techniques; this act could question the ethics of the research. When researchers perform questionable research practices, research integrity is damaged and diminished.

The TTAT instrument was tested by [32], who found that all constructs had a reliability coefficient higher than .70. This result supported the original study, in which they also found reliability coefficients higher than .70 for each construct [10]. In this study, the reliability coefficient was higher than .70, which is consistent with the survey instrument. A consistent result indicates a reliable instrument [59].

# 4. REGRESSION ANALYSIS

This section describes the outcomes of the regression analysis. After examining the linear relationship, normality, multicollinearity, and homoscedasticity of the variables, the study conducted regression analysis for single and multiple independent variables. The results of the regression allowed the analysis of the relationship between the dependent and independent variables. The research question was, to what extent, if any, does perceived severity, perceived susceptibility, perceived threat, safeguard effectiveness, safeguard cost, self-efficacy, avoidance motivation, and avoidance behavior influence mobile device users' susceptibility to phishing attacks? From the research question, nine hypotheses ensued. Below is the description of the analysis of each hypothesis.

Hypothesis 1a. Perceived susceptibility of being attacked by phishing positively affects perceived threat [10]. The analysis of the variance showed the effect of perceived susceptibility of being attacked by phishing on users' perceived threat. A regression was calculated to predict the effect of the perceived threat of being attacked by phishing based on perceived susceptibility. The regression model was significant $F(1, 121) = 69.4$, $p < .001$, with an $R^2$ of .364, a large effect size according to Cohen (1992). The statistical significance is less than .05 which indicates that perceived susceptibility of being attacked by phishing statistically significantly predicted perceived threat. Perceived susceptibility accounted for 36.4% of the variation in the perceived threat of phishing attack with an adjusted $R^2 = 35.9\%$. The significance value was less than .05; therefore, the null hypothesis was rejected.

Hypothesis 1b. Perceived severity of being attacked by phishing positively affects perceived threat [10]. Survey results supported this hypothesis. The dependent variable is perceived threat referring to questions 16-20 on the survey instrument, and the independent variable is perceived severity which refers to question 6-15 on the survey instrument. The analysis of the variance showed the effect of perceived severity of being attacked by phishing. A regression was calculated to predict the effect of the perceived threat of phishing attack based on users' perceived severity of the attack. The regression equation was $F(1, 121) = 379.6$, $p < .001$, with $R^2 = .758$, a large effect size according to Cohen (1992). The statistical significance is less than .05 which indicates that perceived severity accounted for 75.8% of the variance in the perceived threat of being attacked by phishing with an adjusted $R^2 = 75.6\%$. The null hypothesis was rejected.

Hypothesis 1c. Perceived susceptibility and perceived severity of a phishing attack have a positive interaction effect on perceived threat [10]. Survey results supported this hypothesis. The dependent variable is perceived threat which refers to questions 16-20 on the survey instrument and, the independent variables are perceived severity which refers to question 6-15 on the survey instrument and perceived susceptibility which refers to question 1-5 on the survey instrument. Multiple linear regression was calculated to predict the effect of the perceived threat of phishing attack on mobile users' perceived severity and perceived susceptibility of the attack. A significant regression equation was $F(2, 120) = 226.98$, $p < .001$, with $R^2 = .791$, a large size effect according to Cohen (1992). The statistical significance is less than .05 which indicates that perceived severity and perceived susceptibility accounted for 79.1% of the variance in the perceived threat of being attacked by phishing with an adjusted $R^2 = 78.7\%$.

Hypothesis 2. The perceived threat of being phished positively affects avoidance motivation [10]. Survey results supported this hypothesis. The dependent variable, in this case, is avoidance motivation which refers to question 40-42 in the survey instrument and the independent variable is perceived threat referring to questions 16-20 in the survey instrument. The analysis of the variance showed the effect of the perceived threat of being attacked by phishing on users' avoidance motivation. A regression was calculated to predict the effect of avoidance motivation against phishing attacks based on users' perceived threat. The regression equation was $F(1, 121)$ = 132.06, p < .001, with an $R^2$ of .522, a large effect size according to Cohen (1992). The statistical significance is less than .05 which indicates that the perceived threat of being attacked by phishing statistically significantly predicted users' avoidance motivation. Perceived threat accounted for 52.2% of the variation in perceived threat of phishing attack with an adjusted $R^2$ = 51.8%. The significance value was less than .05; therefore, the null hypothesis was rejected.

Hypothesis 3. Safeguard effectiveness against phishing attacks positively affects avoidance motivation [10]. Survey results supported this hypothesis. The dependent variable, in this case, is avoidance motivation which refers to question 40-42 in the survey instrument and the independent variable, is safeguard effectiveness referring to questions 21-26 in the survey instrument. The analysis of the variance showed the effect of users' safeguarding effectiveness and avoidance motivation against phishing attacks. The regression equation was $F(1, 121)$ = 183.26, p < .001, with an $R^2$ of .602, a large effect size according to Cohen (1992). The statistical significance is less than .05 which indicates that avoidance motivation statistically significantly predicted users safeguard effectiveness. Safeguard effectiveness accounted for 60.2% of the variation in users' avoidance motivation of protection against phishing attack with an adjusted $R^2$ = 59.9%. The significance value was less than .05; therefore, the null hypothesis was rejected.

Hypothesis 3a. Perceived threat of phishing attacks and safeguard effectiveness against phishing have a negative interaction effect on avoidance motivation [10]. Survey results supported this hypothesis. The dependent variable was avoidance motivation referring questions 16-20 on the survey instrument, and the independent variables are perceived threat which refers to question 16-20 on the survey instrument and safeguard effectiveness which refers to question 21-26 on the survey instrument. Multiple linear regression was calculated to predict the effect of motivation to avoid phishing attacks on mobile users' perceived threat and safeguard effectiveness. The regression equation was $F(2, 120)$ = 97.10, p < .001, with $R^2$ = .618, a large size effect according to Cohen (1992). The statistical significance is less than .05 which indicates that perceived threat and safeguard effectiveness accounted for 61.8% of the variance in avoidance motivation of being attacked by phishing with an adjusted $R^2$ = 61.2%.

Hypothesis 4. Safeguard cost against phishing attacks negatively affects avoidance motivation [10]. Survey results supported this hypothesis. The dependent variable, in this case, is avoidance motivation which refers to question 40-42 in the survey instrument and the independent variable, is safeguard cost referring to questions 27-29 in the survey instrument. The analysis of the variance showed the effect of users' safeguarding cost and avoidance motivation against phishing attacks. The regression equation was $F(1, 121)$ = 21.21, p < .001, with an $R^2$ of .149, a small effect size according to Cohen (1992). The statistical significance is less than .05 which indicates that avoidance motivation statistically significantly predicted users cost of using safeguard measures. Safeguard cost accounted for only 14.9% of the variation in users' avoidance motivation to protect against phishing attacks with an adjusted $R^2$ = 14.2%. The significance value was less than .05; therefore, the null hypothesis was rejected.

Hypothesis 5. Self-efficacy for taking safeguard measures against phishing attacks positively affects avoidance motivation [10]. Survey results supported this hypothesis. The dependent variable, in this case, is avoidance motivation which refers to question 40-42 in the survey instrument and the independent variable, is safeguard efficacy referring to questions 30-39 in the survey instrument. The analysis of the variance showed the effect of users' self-efficacy for taking safeguard measures and avoidance motivation against phishing attacks. The regression equation was $F(1, 121) = 138.32$, p < .001, with an $R^2$ of .533, a large effect size according to Cohen (1992). The statistical significance is less than .05 which indicates that avoidance motivation statistically significantly predicted users' self-efficacy of taking safeguard measures. Safeguard efficacy accounted for 53.3% of the variation in users' avoidance motivation of protection against phishing attacks with an adjusted $R^2 = 53\%$. The significance value was less than .05; therefore, the null hypothesis was rejected.

Hypothesis 6. Avoidance motivation positively affects the avoidance behavior of using safeguards [10]. Survey results supported this hypothesis. The dependent variable, in this case, is avoidance behavior which refers to question 43-44 in the survey instrument and the independent variable is avoidance motivation referring to questions 40-42 in the survey instrument. The analysis of the variance showed the effect of users' avoidance motivation and avoidance behavior against phishing attacks. The regression equation was $F(1, 121) = 134.40$, p < .001, with an $R^2$ of .526, a large effect size according to Cohen (1992). The statistical significance is less than .05 which indicates that avoidance behavior statistically significantly predicted users' avoidance motivation. Avoidance motivation accounted for 52.6% of the variation in users' avoidance behavior of adopting safeguard measures against phishing attacks with an adjusted $R^2 = 52.2\%$. The significance value was less than .05; therefore, the null hypothesis was rejected. Table 2 provides a summary of the findings for all hypotheses.

Table 2. Summary of findings.

| | Hypothesis | Statistics | Findings |
|---|---|---|---|
| H1a | Mobile device users' perceived susceptibility of being attacked by phishing positively affects their perception of threat. | $F(1, 121) = 69.4$, $p < .001$ | Supported |
| H1b | Mobile device users' perceived severity of being attacked by phishing positively affects their perception of threat. | $F(1, 121) = 379.6$, $p < .001$ | Supported |
| H1c | Perceived susceptibility and perceived severity of a phishing attack have a positive interaction effect on perceived threat. | $F(2, 120) = 226.98$, $p < .001$ | Supported |
| H2 | Perceived threat of being phished positively affects avoidance motivation. | $F(1, 121) = 132.06$, $p < .001$ | Supported |
| H3 | Safeguard effectiveness against phishing attacks positively affects avoidance motivation. | $F(1, 121) = 183.26$, $p < .001$ | Supported |
| H3a | Perceived threat of phishing attack and safeguard effectiveness against phishing has a negative interaction effect on avoidance motivation. | $F(2, 120) = 97.10$, $p < .001$ | Supported |
| H4 | Safeguard cost against phishing attacks negatively affects avoidance motivation. | $F(1, 121) = 21.21$, $p < .001$ | Supported |
| H5 | Self-efficacy for taking safeguard measures against phishing attacks positively affects avoidance motivation. | $F(1, 121) = 138.32$, $p < .001$ | Supported |
| H6 | Avoidance motivation positively affects the avoidance behavior of using safeguard measures. | $F(1, 121) = 134.40$, $p < .001$ | Supported |

Table 3 provides the model summary of the nine dependent variables. This study conducted a regression analysis after confirming the homoscedasticity and normality of the residuals. The study met all regression assumptions. However, the Durbin-Watson value on avoidance motivation had a lower value. Six participants were outliers and were removed from the analysis. Hypothesis 1a, 1b, 1c, 2, 3, 3a, 4, 5, and 6 were all supported based on the statistical testing.

Table 3. Model Summary.

| Variables | $R$ | $R^2$ | Adjusted $R^2$ | Standard Error | Durbin-Watson |
|---|---|---|---|---|---|
| Perceived Susceptibility | .604 | .364 | .359 | 1.279 | 1.588 |
| Perceived Severity | .871 | .758 | .756 | 0.789 | 2.271 |
| Perceived Susceptibility and Severity | .889 | .791 | .787 | 0.737 | 1.868 |
| Perceived Threat | .722 | .522 | .518 | 1.219 | 2.007 |
| Safeguard Effectiveness | .776 | .602 | .599 | 1.112 | 1.779 |
| Safeguard Effectiveness, Perceived Threat | .786 | .618 | .612 | 1.094 | 1.809 |
| Safeguard Cost | .386 | .149 | .142 | 1.626 | 1.658 |
| Self-Efficacy | .730 | .533 | .530 | 1.204 | 1.799 |
| Avoidance Motivation | .725 | .526 | .522 | 1.363 | 1.381 |

# 5. DISCUSSION

The primary aim of the present study was to explore mobile device users' susceptibility to IT threats such as phishing attacks and how motivated users were to avoid such threats. Phishing attacks are a popular means of attaining personal information by cybercriminals for fraudulent use, and it is necessary to understand how to mitigate these attacks most effectively [62]. The study presented here drew on TTAT to investigate how mobile device user behavior impacted the phishing attack landscape. The TTAT is a model used to examine IT threats avoidance and users' behavior given safeguard measures [29].

## 5.1. Implications for Practice

This study examines user IT threat avoidance behavior in the context of mobile devices because of their increased vulnerability to phishing attacks. In 2012, 99% of malware detection targeted Android devices – a form of a mobile device; malware attacks have dominated these devices [63]. In organizations, employees often receive training in security awareness, and they understand their accountability of behavior towards IT security [64]. Security policies are implemented and mandated in organizations, which in turn has attracted more information security research [65]. However, mobile device users are often not mandated by any organizational policies and can be victims of IT threats because they become prey. Anti-phishing applications were developed to thwart phishing and provide awareness to users about phishing attacks [29] but are not sufficient to thwart phishing attacks [66], [6] on mobile devices.

This research indicates that consumers' perception of threat best predicts 52.2% of phishing attack avoidance motivation. As indicated by the results, consumers' perception of threat will increase if they perceive that the consequences of the threat to their mobile device will be severe. The indicated result may encourage users to increase self-efficacy and adopt safeguard measures on their mobile devices. Thus, using safeguard measures on mobile devices could be increased based on their motivation to avoid phishing attacks. To reduce phishing attacks and increase consumer's motivation to avoid the threat, IT security professionals, organizations, and mobile device manufacturers should promote awareness about IT threats using resources such as device manuals, product support, and FAQs. This research is the first study to expand the work of [10] by using phishing attacks as the IT threat and mobile device users in the United States as the populations. Also, this study is the first to evaluate users' susceptibility to phishing attacks on mobile devices using the TTAT. This study provides practitioners with a new understanding of phishing attack avoidance behavior among mobile device users.

## 5.2. Research Limitations and Future Research

Though designed to be generalizable, this study had several limitations. While power analysis indicated that the sample size would be enough to detect statistical significance, this study was limited to a small number of participants that may not understand the meaning of phishing. Before taking the survey, there was no explanation of phishing given to the participants, leaving the possibility that the participants responded to the survey without explicit knowledge of phishing attacks. On the other hand, prior knowledge about phishing attacks could have impacted the data and the analyses of the results.

Another limitation is the length of the survey. The survey consisted of 44 questions which could pose a lengthy process for some participants. It is possible that participants did not take their time answering the survey. Furthermore, the results of the survey could be different if this study used the median score instead of the mean score. The survey instrument used in this survey is reliable, but it could benefit from reconstructing the Likert scales. The study calculated the mean of each section of the survey instrument. Though this study has its limitations, the results were similar to the original research by [10].

## 5.3. Recommendations for Further Study

Studies have shown that mobile device security research is a growing field that continues to attract researchers. Although there are various studies on mobile device security, few have considered different forms of IT threats in mobile device security. Therefore, it is essential for researchers to continue to examine changing IT threats concerning mobile device security. Exploring how individuals utilize their mobile device security features will provide insight into security threats. Understanding how mobile device users implement available security software on their mobile devices is an opportunity for further research.

This study focused on mobile devices such as smartphones and tablets and restricted wearable devices, but a similar study could expand the research to include wearable devices such as smartwatches and wearable fitness monitors. The original research by [10] did not find a correlation between perceived severity and perceived susceptibility concerning users' avoidance motivation. However, this study found a positive correlation; this could be due to increased awareness of IT threats in recent days. Further research, possibly using a larger sample size, is needed to confirm or refute the positive correlation between perceived severity, perceived susceptibility, and avoidance motivation. Further study is required to focus on the age and demographic relations to confirm findings [10]. Because demographic features of the sample, especially age and education level, can impact the level of experience or knowledge of IT and security threats, an additional study that includes these characteristics as variables would be informative. This study could be further modified to include new threats such as hackers and fewer constructs from the TTAT.

## 6. CONCLUSION

Previous research gaps and limitations identified by previous researchers inspired this study. The study utilized the Technology Threat Avoidance Theory questionnaire [10] to test IT threats and users' behavior. The survey instrument contained 44 Likert-Type scale questions. Survey data collected from 137 mobile device users were tested using the survey instrument. Descriptive statistics, frequencies, Pearson correlation, and regression analysis were used to examine the data collected for statistical analysis. The result of the data analysis rejects the null hypothesis for 1a, 1b, 1c, 2, 3, 3a, 4, 5, and 6, while it supports the alternative hypothesis.

The findings from this study provided information about mobile device users' susceptibility to phishing attacks, mobile device users' security practice behavior, and mobile device users' threat avoidance motivation. The study revealed several findings in the analyses: Avoidance motivation determines avoidance behavior of users, perceived susceptibility and perceived severity positively affect perceived threat, and perceived threat strongly determines avoidance behavior. Thirdly, safeguard effectiveness, safeguard cost, and self-efficacy interact with avoidance behavior. Finally, safeguard cost and perceived threat negatively impact a users' motivation to avoid a threat.

This study examined factors that impact mobile device users' susceptibility to phishing attacks in the United States. The research question for this study was "To what extent, if any, does perceived severity, perceived susceptibility, perceived threat, safeguard effectiveness, safeguard cost, self-efficacy, avoidance motivation, and avoidance behavior influence mobile device users' susceptibility to phishing attacks?" Results from the research indicate a positive correlation between perceived threat, avoidance behavior, and avoidance motivation. The study shows that mobile device users feel threatened if they perceive that the severity of the attack will affect them and, therefore, affects their motivation to avoid phishing attacks threat. Mobile device users in the United States might not take appropriate actions to thwart IT threats, thereby making them susceptible to phishing attacks.

Regarding further research (see Section 5), mobile device security provides new ground for exciting and new research. Further study can improve mobile device users' security behavior, discover new types of security threats, and understand the use of mobile device security software. Researchers could shed more insights on mobile users' security threats by applying valuable techniques. Increased mobile device security awareness can help thwart IT threats; however, users must adopt positive behavior to reduce phishing attack threats.

## AUTHOR

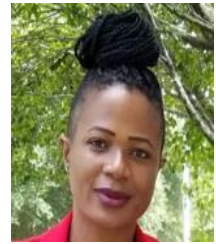

**F. Ley Sylvester** is an Adjunct Professor at Tarrant County College, where she teaches business computers, including computer security practices and other courses. Ley earned a Ph.D. in Information Assurance and Cybersecurity and a Masters in Information Management. Her background includes extensive hands-on experience in information technology and knowledge of various business disciplines. Ley is a team leader and has managed different projects in different sectors, and her abilities include management, leadership, communication, mentoring, and motivation. Her research interests include mobile security, cyber security, and social engineering.